\documentclass[12pt,preprint]{aastex}
\slugcomment{Accepted for publication
in {\it the Astrophysical Journal} and scheduled for 1 December
2004, vol 616 issue} 
\def\lax {\ifmmode{_<\atop^{\sim}}\else{${_<\atop^{\sim}}$}\fi} 
\def\gax {\ifmmode{_>\atop^{\sim}}\else{${_>\atop^{\sim}}$}\fi} 

\begin{document}


\title{ Downscattering due to Wind Outflows in Compact X-ray Sources:
Theory and Interpretation}

\author{Lev Titarchuk\altaffilmark{1,2} and Chris Shrader\altaffilmark{2,3}}

\altaffiltext{1}{George Mason University/Center for Earth
Observing and Space Research, Fairfax, VA 22030; and US Naval Research
Laboratory, Code 7655, Washington, DC 20375-5352; ltitarchuk@ssd5.nrl.navy.mil }
\altaffiltext{2}{NASA/ Goddard Space Flight Center, code 660, Greenbelt 
MD 20771, lev@lheapop.gsfc.nasa.gov, Chris.R.Shrader@gsfc.nasa.gov}
\altaffiltext{3}{Universities Space Research Association, Seabrook MD
20706}

\begin{abstract}
A number of recent lines of evidence point towards the presence
of hot, out-flowing plasma from the central regions of compact 
Galactic and extragalactic X-ray sources. Additionally, it has
long been noted that many of these sources exhibit an ``excess''
continuum component, above $\sim 10$ keV, usually attributed
to Compton Reflection from a static medium. 
Motivated by these facts, as well as by recent observational constraints on
the Compton reflection models -- specifically 
apparently discrepant variability timescales for line and
continuum components in some cases -- we
consider  possible  effects of out-flowing plasma on the
high-energy continuum spectra of accretion powered
compact objects. We present a general formulation for
photon downscattering diffusion which includes 
recoil and Comptonization effects due to 
divergence of the flow.
We then develop  an  analytical theory for the spectral formation
in  such systems that allows us to  derive 
formulae for  the emergent spectrum.
Finally we perform the analytical model 
fitting on several Galactic X-ray
binaries. Objects which have been modeled with 
high-covering-fraction Compton
reflectors, such as GS1353-64 are included in 
our analysis. In addition,
Cyg X-3, is which is widely believed to
be characterized by  dense circumstellar winds 
with temperature of order $10^6$ K,
provides an interesting test case. Data from INTEGRAL and
RXTE covering the $\sim 3-300$~keV range are used in our analysis. 
We further consider the possibility that the widely noted distortion of the power-law
continuum above 10 keV may in some cases be explained
by these spectral softening effects.
\end{abstract}


\section{INTRODUCTION}
Recent observational and theoretical evidence suggest that accretion-powered
X-ray sources, of both the Galactic and extragalactic variety, may 
exhibit outflowing plasma, i.e. winds,  emanating from a compact 
region near the central source
(e.g. Elvis 2003; Arav 2003; Brandt \& Schulz 2000; Proga \& Kallman 2002).
Comptonization effects in those putative outflows  are 
likely to alter the intrinsic continuum 
spectra of accretion powered compact objects. The basic idea is that
electron  scattering of photons from a central source
entering the expanding outflow experience a decrease in
energy (downscattering). The magnitude of this decrease is of first order 
in $v/c$ and in $E/m_ec^2$ where $v$ is the outflow speed, 
c is the speed of light, $E$ is the initial photon energy, 
and $m_e$ is the electron rest mass. 

 The basic idea is depicted in Figure 1. There  present a simple explanation 
of the diverging flow  effect 
on the photon propagation through the medium. 
A photon emitted outwards near inner boundary and then 
scattered at a certain point by an electron moving with velocity ${\bf v}_1$, is
received by an electron moving with velocity ${\bf v}_2$ as shown with
frequency $\nu _2 = \nu _1\left[1+\left({\bf v}_1-{\bf v}_2\right)
\cdot{\bf n}/c\right]$ where ${\bf n}$ is a unit vector along the
path of the photon at the scattering point.  In a diverging flow
$\left({\bf v}_1-{\bf v}_2\right)\cdot{\bf n}/c <0$ and photons are
successively redshifted, until scattered to an observer at infinity.
 The color of photon path  (in Figure 1) indicates
the frequency shift in the rest frame of the receiver (electron or the Earth
observer). On the other hand, referring to the right-hand side of Figure 1, 
in a converging flow
$\left({\bf v}_1-{\bf v}_2\right)\cdot{\bf n}/c >0$ and photons are
blueshifted.

The classical Compton (recoil) effect $<\Delta E>/E \sim -E/m_ec^2$ has been well understood 
for a long time. 
Basko, Sunyaev \& Titarchuk (1974), hereafter BST74 first studied the downscattering effects
in the interaction of X-ray radiation of the central source 
with the relatively cold atmosphere of the optical companion in a binary system.
They predicted the shape of the X-ray  reflection spectrum of the companion and applied
these results to the Her X-1 system.  X-ray observations of Her X-1 by a number of groups 
(e.g. Sheffer et al. 1992, Still et al. 2000 and etc.) confirmed this  prediction.
In particular, the numerical calculations by BST74 demonstrated that the reflection 
of the bremsstrahlung spectrum distorted the high-energy 
continuum above 10 keV producing the  characteristic continuum excess feature, or ``bump'', 
in the $>10$ keV spectrum. This feature 
was also identified  by Sunyaev \& Titarchuk (1980), hereafter ST80, who found
that the transmitted, downscattered spectrum is formed as a result 
of the reprocessing of X-ray radiation from the central source in an ambient 
spherical cloud.
 From these facts  one can conclude that this distortion of the 
continuum is not an intrinsic property
of the particular space photon distribution, but it is
rather a result of {\it effective} downscattering (in terms of the mean number of scattering and the energy loss per scattering) 
 X-ray photons  diffusing through  the intervening medium.

In this paper we show that this is  precisely the case. We formulate the 
problem of the  photon diffusion in generic terms and 
demonstrate a specific solution for bulk 
motion where effects of a recoil and divergence of the flow are taken into account. 
The emergent spectrum as an outcome of this solution has all
these aforementioned features of the downscattering.

The details of the radiative transfer problem taking into account downscattering effects 
(recoil and Doppler effect in the divergent flow)  and its solution are   given in \S 2.
In section 3, we apply our model to observational data for several Galactic X-ray binaries. 
We attempt to demonstrate
the Compton downscattering effects in relatively cold outflows for 
which temperature is of order 
$10^6$ K and  disk atmospheres which are necessary constituents of X-ray
sources. For example, Cygnus X-3 is believed to be characterized by  dense circum-stellar
(or circum-disk) winds, provides an interesting test case.
A number of additional sources, which have been noted in the literature
to exhibit evidence for strong``Compton reflection'', e.g. GS1353-64 and GX339-4, 
have also been  included in our analysis. We consider the possibility that this well
documented distortion of the power-law continuum above
10 keV, may in some cases be due to downscattering from
out-flowing plasma rather than from a static reflecting
media. We further speculate that this  could possibly apply to
extra-galactic (i.e. AGN) as well as Galactic accretion
objects (\S 4), although we do not present any analysis of AGN here.
 A summary and conclusions follow in \S 5.

\section{RADIATIVE TRANSFER IN A BULK OUTFLOW}

\subsection{The Basic Downscattering Problem}

The problem of photon propagation in a fluid in bulk motion has been studied in
detail in a number of papers [see e.g.  Blandford \& Payne (1981), hereafter
BP81; Payne \& Blandford (1981), hereafter PB81; Nobili, Turolla
\& Zampieri (1993); Titarchuk, Mastichiadis \& Kylafis (1997); Titarchuk,
Kazanas \& Becker (2003, hereafter TKB03); Laurent \& Titarchuk (1999), (2001),
 (2004, hereafter LT04].
In particular, TKB03 present a general formulation and a solution of the spectral
formation in the diverging outflow. They demonstrated
that the resulting spectrum can be formed as a convolution of energy and space diffusion
solutions.  The spread function, given by a  Green's function formulation, 
applied to a monochromatic injection yields a redshift-skewed line which
is a power law at low energies. Furthermore, Monte-Carlo simulations show that 
the line width has a strong dependence on $v/c$, 
optical depth $\tau$ and the energy of the monochromatic line
$E_0$ only when the flow temperature $kT_e<1$ keV (LT04).  
TKB03 and LT04 also establish that the shape of the spectral line  should be closely related
to the photon source distribution in the flow. Comparison of the theoretical spectra with data
show that observed red-skewed lines are formed as a result of transmission of the X-ray radiation
through outflows of moderate Thomson optical depth ($\tau_0 \gax 1$).

To apply these Radiative Transfer results to observational data,
we have developed a generic analytical formulation which leads to a simple
analytic expression for modification of the emergent 
spectrum due to recoil and velocity 
divergence effects in the flow.  For the recoil effect, we extend 
the  results of ST80
 who show that the downscattering feature, or "bump", can
appear superposed on power-law spectra with energy spectral indices $\alpha<1$  
as a result of photon
diffusion through a static cloud.  From ST80 and  calculations presented here,
one might conclude such bumps are not a necessarily a feature of
disk reflection, but could instead be a generic feature of the photon 
reprocessing  in a relatively cool, ambient plasma characterized by 
temperatures of order $10^6$ K.

\subsection{Main Equations and  Solution of Downscattering Problem }

Let $N(r)=N_0(r_0/r)^{b}$ be the radial number density profile of an
outflow and let its radial outward speed be

\begin{equation}
v_b/c=(\dot M_{out}/4\pi cm_pN_0r_0^2)(r_0/r)^{2-b}=
(v_0/c) (r_0/r)^{2-b}
\end{equation}
obtained from mass conservation in a spherical geometry (here $\dot M_{out}=4\pi r^2v_bm_pN$ and $m_p$
is the proton mass). 
The Thomson optical depth of the flow from some radius r to infinity is given by

\begin{equation}
\tau=\int_r^{\infty}N_e(r)\sigma_{\rm T}~dr=\sigma_{\rm T}N_0r_0(r_0/r)^{b-1}/(b-1),
\end{equation}
where $N_e(r)=N(r)$ is the electron density, $\sigma_{\rm T}$ is the Thomson cross 
section, $r_0$ is a radius at the base of the outflow. Because our final results
are independent of the velocity and density profiles  below  for simplicity of
presentation we use  $b=2$ (a constant velocity outflow).  
In this case  $\tau_{0}=\tau(r_0)=\sigma_{\rm T}N_0r_0$.


The transfer of radiation within the flow in space and energy is governed by 
the photon kinetic equation (BP81, Eq. 18) for the photon occupation
number $n(r,\nu)$, which in steady state reads
%
%
\begin{equation}
-\frac{{\bf v_b}/c}{\kappa}\cdot\nabla n+\frac{1}{3\kappa}\nabla\cdot(\frac{1}{\kappa}\nabla n)
+\frac{1}{3}\frac{\nabla\cdot{({\bf v_b}/c})}{\kappa}\,
z \frac{\partial n}{\partial z}+\frac{1}{z^2}\frac{\partial(z^4n)}{\partial z}=-{j}(r,\nu),
\label{radtrans}
\end{equation}
where  $z=h\nu/m_ec^2$ is the dimensionless photon energy, $\kappa=N_e(r)\sigma_{\rm T}$ is the inverse of 
the scattering mean free  path, ${\bf v_b}=v_b{\bf e_r}$, is the flow velocity,  ${\bf e_r}$ is the radial 
unit vector and $ j(r,\nu)$ is the photon source term. 

The spectral flux $F(r,\nu)$ (PB81) is given in terms of $n(r,\nu)$ by 

\begin{equation}
F(r,\nu)= -\frac{1}{3\kappa(r)}\nabla n-\frac{1}{3}v_b \, \nu
\frac{\partial n}{\partial \nu},
\label{flux}
\end{equation}
and must satisfy the following boundary conditions: (a) Conservation of 
the frequency integrated flux over the outer boundary, namely 

\begin{equation}
L^{(1)}_r n=\int F(r,\nu)d\nu\propto r^{-2} ~~~{\rm as}~~ {\rm r\rightarrow \infty}.
\end{equation}
(b) 
In terms of the specific intensity of the radiation, the second inner boundary  condition specifies 
that the escaping radiation from the inner boundary be equal to the radiation entering the wind shell 
through the inner boundary (see Fig. 1 for the wind geometry). We neglect here the possible absorption 
of the radiation at the central
source (black hole).  Thus the occupation number should satisfy some kind of the reflection condition
 at the inner boundary $r=r_0$. Namely the  photon flux at the inner boundary is zero 
 (all photons scattered in  the outflow shell and escape through the inner boundary subsequently return).
\begin{equation}
L_r^{(2)}n(r_0)=0.
\end{equation}
In this formulation of the problem the number of photons emitted in the wind shell  equals to that escaped to the Earth observer  (i.e. the photon number is conserved). This can be proven using Eqs. (3), (4-5).
Consequently,  if   the high energy photons lose their energy in the way out but the number of photons is conserved 
{\it it has to be the accumulation of the photons at a particular lower energy band}. 
This accumulation effect was previously noted by ST80 for the case when  the photon energy loss (downscattering) 
was determined by the recoil effect only  (see  Fig. 10 in there).   
In section 2.3 we demonstrate this accumulation effect when the recoil and flow divergence dowscattering
effects are taken into account.
   
It is worth noting that in TKB03 the goal was to demonstrate the pure
effect of the flow divergence on a spectral line as a result of
multiple scatterings with the plasma  free electrons. They thus
omitted the recoil term in the left hand side of equation
(\ref{radtrans}).  They were also interested to applying their solution to
the  spectral  formation  of iron  lines for which photon energies are
about 6.4 keV and lower.  For such energies the recoil effect can
be safely neglected (see ST80). On the other hand if one is
interested in the modification of continuum spectrum  at energies
higher than 10 keV by multiple downscattering events in the ``cold'' medium
this effect must be taken into account.   Below  we show how one
can generalize  the TKB03 solution including the recoil term  in the
radiative transfer equation.  We apply  a particular method for the
separation of variables suggested in Titarchuk (1994), hereafter T94,  and
successfully used in TKB03, to obtain our solution.

In order to further proceed with the solution derivation we should note that 
Laurent and Titarchuk (2004) find  that the coefficient in the diverging term of equation (\ref{radtrans})
 $\tilde\varepsilon=\nabla\cdot({\bf v_b}/c)/\kappa$ can be replaced by a constant $\tilde\varepsilon$. 
Strictly speaking $\tilde \varepsilon$  is a function of $r$ and consequently of
$\tau$, namely $\tilde \varepsilon=2(v_b/c)/\tau$.
  
 Using the Fokker-Planck equation [see Eq. (3) in TKB03] and a method
developed  by Titarchuk, Mastichiadis \& Kylafis (1997) (see Appendix
D there)  one can relate the  mean energy change per scattering,
$<\Delta E>$,  for photons undergoing numerous scatterings in the flow
to $\beta=v/c$:  
\begin{equation}
<\Delta E>\approx-[f{\bf \nabla v}/(c\kappa)]E\approx -(2f\beta/\tau_{0}) E\label{deltaE}
\end{equation}
where ${\bf v}=v{\bf e_r}$ is the flow velocity, ${\bf e_r}$ is the
radial unit vector, $\kappa=N_e\sigma_{\rm T}$ is the inverse of
scattering mean-free-path $l$.  The numerical factor $f$,  in formula
(\ref{deltaE}) is of order unity and LT04  obtain its precise value
using their MC simulations.

In LT04  (specifically, Figure 2 of that paper) the escape photon
distribution $\varphi(t)$ for 5  energy bands is present.  In  the
plot the time is given in light crossing time units $t_{cross}=\Delta
r/c$  where $\Delta r$ is the outflow cloud thickness.  Photons which
escape without any scattering are at 6.6 keV.  The model parameters
$kT_e = 0.1$ keV, $\tau_0 =4, ~\beta=v/c = 0.1$. 

The simulated  time distribution  is fitted  by an exponential
$\varphi(t)=C_N\exp(-at/t_{cross})$,  with $a$ equal to 0.67
identical to that predicted by diffusion theory [see Sunyaev \&
Titarchuk  in 1985 (hereafter ST85)].  ST85 show that the average
number of scatterings $N_{av}$ in the shell of  optical depth $\tau_0$
is $3\tau_0^2/8$ so the average photon scattering  time is
$t_{av}=N_{av}l/c=3\tau_0 t_{cross}/8$, 
where $l=1/N_0\sigma_{T}=\Delta r/\tau_0$.  
ST85 also show that the time distribution for  scattered
photons in any bounded medium is an exponential, $\varphi
(t)=C_N\exp(-t/t_{av})$  where $C_N=1/t_{av}$ ($\int_0^{\infty}f(t)dt=1$). 

In this particular case 
$f(t)=C_N\exp(-t/t_{av})=C_N\exp[-(2/3)/(\tau_0/4)(t/t_{cross})]$,
precisely what is obtained in the LT04 simulations.  Thus,  the
average energy of photons escaping after $N_{av}-$scatterings is
$<E>_{sc}\approx(1+<\Delta E>)^{N_{av}}$.  Using formula
(\ref{deltaE})   one can obtain that 
\begin{equation}
<E>_{sc}\approx(1-2f\beta/\tau_0)^{N_{av}} E_0.
\label{esc}
\end{equation}
For the particular case of $\tau_0=4$  and $\beta=0.1$ 
LT04  find that the original photon
 energy $E_0=6.6$ keV is reduced  to $E_0=5.4$ keV, after
 $N_{av}-$scatterings  in the outflow [the emergent spectrum is shown
 in Fig 1, (LT04) left panel].  The analytical estimate obtained from
 formula (\ref{esc}) is close to   this value of  $<E>_{sc}=5.4$ keV
 for $f\approx2/3$.  Thus  one considers an approximation in which the
 diverging term in equation (\ref{radtrans}) is independent of space
 variable. 

\subsection{Downscattering solution: Emergent spectrum} 
According to a theorem (T94, appendix A), the solution of any 
equation whose LHS operator acting on the unknown function  $n(r,\nu)$
is the sum of two operators $L_{r}$ and $L_{\nu}$, which depend
correspondingly only on space and energy and the RHS, $j(r,\nu)$,
can be factorized, i.e. 
\begin{equation}
L_rn+L_{\nu}n=-j(r,\nu)=-f(r)\varphi(\nu).
\label{maineq}
\end{equation} 
The boundary conditions independent of the energy $\nu$, i.e.
\begin{equation}
L^{(1)}_rn=0~~~~{\rm as}~~{\rm r\rightarrow \infty}, ~~~~ 
L^{(2)}_rn=0~~~~{\rm for}~~{\rm r=r_0}~, 
\label{boundeq}
\end{equation}
$n$ is given by the convolution of the solutions of the time-dependent problem 
of each operator, namely
\begin{equation}
n(r,\nu)=\int_0^{\infty}P(r,u)X(\nu,u)du. 
\label{convol}
\end{equation}
Above $u$  is the dimensionless time (which can be the Thomson dimesionless time $u_{\rm T}=N_0\sigma_{\rm T}ct$ depending on the
specific forms of operators $L_r$ and $L_{\nu}$) and 
$P(r,u)$  is the solution of the initial value problem of the 
spatial operator $L_{r}$
\begin{equation}
\frac{\partial P}{\partial u}=L_rP, ~~~~~P(r,0)=f(r)
\label{spaceq}
\end{equation}
with boundary conditions
\begin{equation}
L^{(1)}_rP=0~~~~{\rm as}~~{\rm r\rightarrow \infty},~~~~
L^{(2)}_rP=0~~~~{\rm at}~~{\rm r=r_0},
\label{spacebc}
\end{equation}
%
and $X(\nu,u)$ the solution of the initial value problem of the energy
operator  $L_{\nu}$
\begin{equation}
\frac{\partial X}{\partial u}=L_{\nu}X, ~~~~~X(z,0)=\varphi(z)
\label{energyeq}
\end{equation}
%
with  boundary conditions
\begin{equation}
z^3X\rightarrow 0~~~~{\rm when}~~~z\rightarrow 0,~\infty.
\label{energybc}
\end{equation}
Thus we have 
\begin{equation}
L_rP=-\frac{{\bf v_b}/c}{\kappa}\cdot\nabla P+ \frac{1}{3\kappa}\nabla\cdot(\frac{1}{\kappa}\nabla P).
\end{equation}

 Following, the LT04 arguments 
we replace the coefficient of the diverging term  by a constant $\varepsilon=2q(v_b/c)/\tau_{\rm T, 0}\ll1 $ where a 
numerical factor $q$ is order of unity (see details in the end of \S 2.2). 

In the case where
the photon energy is due to a recoil effect and Comptonization effects in the
diverging flow, we have 
\begin{equation}
\frac{\partial X}{\partial u}=L_{\nu}X=\frac{1}{3}\varepsilon z \frac{\partial X}{\partial z}+
\frac{1}{z^2}\frac{\partial(z^4X)}{\partial z}.
\label{energyeq2}
\end{equation}
\begin{equation}
X(z,0)=\varphi(z)/z^3
\label{energyeqin}
\end{equation}
with  boundary conditions
\begin{equation}
z^3X\rightarrow 0~~~~{\rm when}~~~z\rightarrow 0,~\infty.
\label{energybc2}
\end{equation}
We transform Eq.(\ref{energyeq2}) introducing a new unknown function $Y=e^{(4\varepsilon/3)u}z^4X$ 
for which equation has a form 
\begin{equation}
\frac{\partial Y}{\partial u}=\left(\frac{\varepsilon}{3} z+z^2\right) \frac{\partial Y}{\partial z}.
\label{energyeq2m}
\end{equation}

 The problem for equation (\ref{energyeq2m}) with appropriate initial condition $Y(z,0)=z\varphi(z)$ and
 boundary conditions $Y\rightarrow 0$  when $z\rightarrow 0,~\infty$  is an initial value problem 
for the first order partial differential equation and  it can be found 
using the method of characteristics (see e.g. TKB03).
The differential equation for the characteristics is 
\begin{equation}
du=-\frac{dz}{(\varepsilon/3)z+z^2}
\label{deqcharact}
\end{equation}
that solution is 
\begin{equation}
 u=\frac{3}{\varepsilon}[\ln (z+\varepsilon/3)/z-\ln(z_0+\varepsilon/3)/z_0)],
 \label{charact}
\end{equation}
where $z_0$ is the dimensionless energy at $u=0$.
Because  $Y(z_0,0)=z_0\varphi(z_0)$ is conserved along the characteristics
 the solution of the problem (\ref{energyeq2}-\ref{energybc2}) is
\begin{equation}
J(z,u)=z^3X(z,u)=e^{-4\varepsilon u/3}\varphi[\psi^{(\varepsilon)}(z,u)]/[z\psi^{(\varepsilon)}(z,u)],
\label{Jnuu}
\end{equation} 
where $z_0=\psi^{(\varepsilon)}(z,u)=(\varepsilon/3)/[(1+\varepsilon/3z)\exp(-\varepsilon u/3)-1]$ is found from Eq.
(\ref{charact}).
Substitution of $J(z,u)$ from Eq.(\ref{Jnuu}) into Eq.(\ref{convol})
gives us the   emergent spectral shape
\begin{equation}
{\cal F}_E(z,\varepsilon)=[\tau^{-2}F(\tau,z)]|_{\tau\rightarrow0}\propto\frac{1}{z}
\int_0^{u_{max, \varepsilon}(z)}e^{-4\varepsilon u/3}
[\psi^{(\varepsilon)} (z,u)]^{-1}
\varphi[\psi^{(\varepsilon)}(z,u)] {\cal P}(u)du 
\label{emerg}
\end{equation}
where $u_{max, \varepsilon}(z)=(3/\varepsilon)\ln(1+\varepsilon/3z)\gg1$ and 
${\cal P}(u)\propto [\tau^{-2}\partial P/\partial\tau[\tau, u]|_{\tau\rightarrow0}$
using the expressions for $F(\tau,z)$  and $P(\tau,u)$  
[see Eq. (\ref{flux}),  Eqs. (\ref{spaceq}-\ref{spacebc}) for the 
definition of $F(\tau,z)$ and $P(\tau,u)$ respectively].

This result is a generalization of ST80 and TKB03 results. ST80 derived the 
spectra when the downscattering effects 
due to the recoil was taken into account. On the other  hand TKB03 considered 
the downscattered spectra as a result of the divergence of the flow.
For example, one can obtain the ST80 formula (36) for the recoil spectrum 
assuming $\varepsilon\to 0$ in 
Eq. (\ref{emerg}). In fact, 
\[ \lim_{\varepsilon \rightarrow0}\psi^{(\varepsilon)}(z,u)=(1/z-u)^{-1}\] 
and $u_{max,0}(z)=1/z$. 

{\it That  formula (\ref{emerg})  for  $\varepsilon=0$ (the recoil case) is
generic and it is valid for any geometric configuration of the plasma
cloud, e.g. a disk, or a spherical cloud, as well as for any photon source
distribution within the cloud, e.g. uniform, or central illumination
distributions.} In fact, what we use here to derive this
formula are  the particular  properties of the diffusion operator of
the left hand side of the main equation (\ref{maineq}),  the boundary
conditions (\ref{boundeq}) and the source term in right hand side of
equation (\ref{maineq}). Namely i. the diffusion operator should be a
sum of two operators, one is the space diffusion operator and another
one is  the energy operator, ii. the boundary condition are
independent of energy and iii. the source function is factorized [or
presented as a linear superposition of the products of
$f(r)\varphi(\nu)$].  All these mathematical properties are generic for
the diffusion problem   and independent of any specific geometry and
spectral  and space source distribution (see ST80 and T94 for more
details and particular examples).

Below we demonstrate how formula (\ref{emerg}) can be simplified 
by exploiting the fact that
 downscattering energy change $\Delta E/E$ is 
proportional to $\varepsilon$ and $E/m_ec^2$. 
 
It should be noted that the downscattering modification of the spectrum occurs
when photons undergo multiple scatterings $u$. It is always on the order of the average
number of scatterings 
$\tilde N_{av}=\int_0^\infty u{\cal P}(u)du/\int_0^\infty {\cal P}(u)du$. 
Thus  we  can expand the integrand function ~~~ 
$W^{(\varepsilon)}(z,u)=e^{-4\varepsilon u/3}[\psi^{(\varepsilon)} (z,u)]^{-1}\varphi[\psi^{(\varepsilon)}(z,u)]$ over $u$
in formula (\ref{emerg}): 
 \begin{equation}
W^{(\varepsilon)}(z,u)\approx W^{(\varepsilon)}(z,0)+W^{(\varepsilon)\prime}_{u}(z,0)u.
 \label{expan}
\end{equation}
This expansion is valid because
  we consider the case when $N_{av}\ll 1/\varepsilon$ ($\tau_{T,0}\gax 1, ~\varepsilon \ll 1$).
Substitution of Eq (\ref{expan}) into Eq. (\ref{emerg}) gives us 
\begin{equation}
{\cal F}(\nu)\propto  z^{-1}[W^{(\varepsilon)}(z,0)+N_{av}W^{(\varepsilon)\prime}_{u}(z,0)],
 \label{intexpan}
\end{equation}
where $z^{-1}W^{(\varepsilon)}(z,0)$ is the incident spectrum $\varphi(z)$ 
and 
\begin{equation}
W^{(\varepsilon)\prime}_{u}(z,0)]/z=\varphi(z)
\left\{\{\ln [z_0\varphi(z_0)]\}_{z_0}^{\prime}|_{u=0}
\frac{\partial z_0}{\partial u}|_{u=0}-4\varepsilon/3\right\},
 \label{deriv}
\end{equation}
where $z_0=\psi^{(\varepsilon)}(z,u)=(\varepsilon/3)/[(1+\varepsilon/3z)\exp(-\varepsilon u/3)-1]$ and
\begin{equation}
\frac{\partial z_0}{\partial u}|_{u=0}=z^2(1+\frac{\varepsilon}{3z}).
 \label{parti}
\end{equation}
Substitution of equations (\ref{deriv}, \ref{parti}) and
$\varphi(z)=z^{-1}W^{(\varepsilon)}(z,0)$ into (\ref{intexpan}) leads us to the
formula:
\begin{equation}
{\cal F}_E(z,\varepsilon)\propto  \varphi(z)\{1-(4\varepsilon/3)\tilde N_{av}+\tilde N_{av}z^2(1+\varepsilon/3z)
[\ln (z_0\varphi(z_0))]_{z_0}^{\prime}|_{u=0}\}.
 \label{intexpanm}
\end{equation}
The most interesting case is the downscattering modification of Comptonization spectrum (see e.g. ST80 and Titarchuk
1994) that can be well fitted by a powerlaw spectrum with an exponential cutoff, namely
 \begin{equation}
\varphi_{comp}(z) \approx z^{-\alpha}\exp(-z/z_{\ast}).
 \label{comp}
\end{equation}
The cutoff energy $E_{\ast}$ is related to the Compton cloud electron temperature $kT_e$,
namely $E_{\ast}\approx 2kT_e$. 
For $\varphi_{comp}(z)$, it is evident that 
\begin{equation}
[\ln (z_0\varphi(z_0))]_{z_0}^{\prime}|_{u=0}= (1-\alpha)/z-1/z_{\ast}
 \label{lnderiv}
\end{equation}
Using equation (\ref{lnderiv}) we transform formula (\ref{intexpanm}) as follows
\begin{equation}
{\cal F}_E(z,\varepsilon)\propto  \varphi_{comp}(z)\{1-(4\varepsilon/3){\tilde N}_{av}+{\tilde N}_{av}z(1+\varepsilon/3z)
[(1-\alpha)-z/z_{\ast}]\}.
 \label{fcomp}
\end{equation}
The modification of the absolute normalization of the spectrum due to
the downscattering effects  is quite obvious $-$ the relative change
of the normalization is $1-(1+\alpha)\varepsilon \tilde N_{av}/3$ .
Thus  we can rewrite  Eq. (\ref{fcomp}) as {\it the final formula of
the spectral shape} as follows 
\begin{equation} 
{\cal F}_E(z,\varepsilon) \propto \varphi_{comp}(z)\{1+N_{av}z[(1-\alpha)-\varepsilon/3z_{\ast}-z /z_{\ast}],  
\label{fcompm}
\end{equation}
where $ N_{av}=\tilde N_{av}/[1-(1+\alpha)\varepsilon \tilde
N_{av}/3]$.  The second term in parenthesis of formula (\ref{fcompm})
describes the pile up and softening of the Comptonization spectrum
$\varphi_{comp}(z)$  due to the downscattering effect in the outflow.
In Figure 2a  we present $E{\cal F}_{E}$ diagram for various $N_{av}$
and $\varepsilon$.  The downscattering (accumulation) bump and softening of spectrum at high energies
are clearly seen in this plot [see also Fig. 2 where a ratio of the Comptonization 
models to the incident spectrum  (\ref{comp})  is plotted].

Below we apply formula (\ref{fcompm}) to the X-ray spectral data for a
number of sources.  The self-consistency of application of this
formula for data fitting can  be checked by comparison of the best-fit
parameter  $N_{av}$ and $u_{max,\varepsilon}(z_{max})$  [the formula
for $u_{max,\varepsilon}(z)$ is shown just after Eq. (\ref{emerg})].
Our inferred best-fit parameter $N_{av}\ll
u_{max,\varepsilon}(z_{max})$ for all fits and thus our formula
(\ref{fcompm})   as the analytical  approximation of (\ref{emerg}) is
valid for all cases considered.

\section{DATA ANALYSIS}

To test these ideas we obtained public data from the INTEGRAL and
HEASARC (RXTE) archives for a sample of Galactic X-ray binaries (Table
1).  The Cygnus region was covered extensively by INTEGRAL  in the
early stages of that mission, Performance Verification Phase,
(November-December 2002),  subsets of which are in the public domain
at the  time of this submission. Specifically, we selected coverage of
Cyg X-3  and Cyg X-1 obtained during revolutions 22-25. Statistics are
limited by the small amount of useful data, given the early problems
with data gaps.   We extracted spectra from JEM-X and SPI, using the
OSA release version 3 software. Subsequent model fitting was done with
the XSPEC spectral analysis package  (version 12.10$\alpha$) modified
to include a local implementation of the model described in section 2.
We also analyzed the JEM-X and SPI data for Cyg X-1, applying the same
model. For  comparison, we also obtained and analyzed contemporaneous
RXTE data for both sources.  In all cases, we used the RXTE
calibration database and software release current as of April, 2004.
We also utilized RXTE data for the additional sources GX 339-4 and GS
1354-63, each of which have been modeled by various other groups --
e.g. Gilfanov et al (1999);  Nowak, Wilms \& Dove (2002);  Pottschmidt
et al (2003); Vilhu et al (2003)  -- as a power-law- exponential plus
a Compton "reflection" continuum (see e.g. Magdziarz \& Zdziarski
1995).  For example, GS 1353-64 was found to require a large
reflection continuum, $R\sim 0.3-0.6$ (Gilfanov et al 2003).  In
addition to the Comptonized continuum, iron line structure - a
gaussian line component and absorption edge feature -  was found to
improve the quality of fit  at low energies. In addition, a systematic
error component of $1\%$ was added to the PCA data. We note that there
is a significant cross-calibration discrepancy between the INTEGRAL
SPI and JEM-X instruments (e.g. Paizis et al 2003).  We have assumed
that the absolute flux calibration of SPI is reliable (Attie' et
al. 2003; Sturner et al. 2003),  and renormalized the JEM-X model fits
accordingly.

In Table 1 we list our source sample, and the inferred parameters of
the out-flow-Comptonization from our analysis.  Here $\Gamma=\alpha+1$
is  the (photon) spectral index.  $N_{av}$ is the average number of
scatterings experienced by a typical photon in the outflow (section
2).  The origin of the data RXTE (PCA plus HEXTE) or INTEGRAL (JEM-X
plus SPI) are also indicated. We note that two  Cyg X-3 observations
represent very different intensity states  for that source. For that
matter, the GX~339-4 and GS~1353-64  observations used correspond to
high-intensity states for each of those sources (but in case, the
spectral  energy distributions represent the low-hard state).

Figure 3 further illustrates the downscattering effect. 
Plotted there is
our best fit model curve, using GX339-4 as a test case, compared to a
simple exponentially-folded power-law fitted to the same data,
with an accompanying residual plot.
This  illustrates 
the improvement in the fit above 10 keV.  It
is clear from these plot that the downscattering effects lead to an
improved fit above $\sim 10$ keV.  Also one can see the photoabsorption features below 10 keV and higher K-edge
(in 7-9 keV energy band) in the residual plot  (Fig. 3a). It is worth the similar edge features along with the strong
$K_{\alpha}$ lines are detected during X-ray superbursts (see Figs. 5 and 9 in Strohmayer \& Brown 2002).
There is a  high probability that they are originated in the radiation driven
outflows during burst events. 

In Figure 4, we show an example of
one of our model fits, in this case to GS~1353-64 (Figure 5 is the
same result, but plotted in photon space).

\section{DISCUSSION}

We have shown  that downscattering modification of the primary photon spectrum
by an outflowing plasma is a possible mechanism 
for producing  the continuum excess in the $\sim10$ keV spectral region. 
This is usually attributed to Comptonization by a static reflector, such as a downward- or
obliquely-illuminated accretion disk,  although the overall continuum
form differs from that of the basic Compton reflection form. We thus
suggest, that  in at least some cases, the outflow downscattering
effect rather than the standard Compton reflection  mechanism is
responsible for the observed ''excess`` hard-X-ray continuum.

It is reasonable to expect that outflowing plasma, in the form of
stellar or putative disk winds will effect the emergent spectra of
compact binaries. Collimated outflows are well known in certain
objects, the so called Galactic micro-quasars, and there is strong
un-collimated or weakly-collimated outflow in additional objects; as
noted Cyg X-3, where in fact spatially extended emission has been
resolved (Heindl et al.  2003), and of course the stellar winds in Cyg
X-1 have been extensively studied. Disk winds may have been observed
directly in Circinus X-1, e.g. Brandt \& Schulz (2000).  Recently,
evidence for disk winds in AGN has emerged (e.g. Elvis 2001; Arav
2003). Further observational confirmation is needed, but if present,
these winds could produce the downscattering effects we propose. 

In addition to evidence for outflows, recent observations suggest that
the putative reflector is,  in at least  certain objects, necessarily
much larger than any reasonable  disk size(e.g. Mattson \& Weaver
2004; Markowitz, Edelson \& Vaughan, 2003).  These arguments are based
on temporal signatures; specifically the lack of a prompt response of
the "reflected" emission to the continuum assuming light travel times
within reasonable accretion disk spatial scales. While an accretion
disk torus has been suggested as an alternative, i.e. more remote,
reflector, the same temporal signatures could be reproduced within the
context of a central source and an ambient outflow. 
          
We note the concerns of some authors, e.g. Miller et al. (2004),  that
the hot disk inner-disk  component and high frequency QPOs (~few times
100 Hz) seen in accreting Black Hole (BH) at the highest inferred
accretion rates  would not be visible through an ``optically thick''
outflow.  Those authors thus rule out any possibility of  ambient
spectral reprocessing and distortion of the iron line feature, such as
broadening and red-wing enhancement, by the outflow. On the other
hand, Laurent \& Titarchuk (2004) infer the outflow Thomson optical
depth $\tau_{0}$ of the out-flowing  medium from parameter fitting
using  XMM  and ASCA measurements of broad, red-shifted iron
lines. They find that $\tau_{0}$ never exceeds ~2-3 in any of the
cases analyzed.  Presumably, the observer sees the radiation of the BH
central source through  the "haze" of the moderate optical
depth. Furthermore Titarchuk, Cui \& Wood (2002), hereafter TCW02
give a precise model for the loss of the modulation due photon
scattering.  It follows from TCW02 that the outflow optical  depth
$\tau_{T,0}$  would need to be around 16 and higher in order to
suppress  QPO amplitude   of frequency 100 Hz [see formula  (5) in
TCW02]. We have shown that this  very optically thick outflow is
ruled out by observations.  Our results confirm the LT04 results  in
the sense that all effects of reprocessing and line  distortion can
plausibly occur in outflows  characterized by moderate optical depth.

Furthermore,  recently Laming \& Titarchuk (2004), hereafter LaT04,  formulate a generic problem of the outflow illumination 
by the hard radiation of the central object and they calculate the outflow temperature and 
ionization balance as a function of the ionization parameter.
Natural assumptions regarding the X-ray spectral distribution of the central source radiation (Comptonization like spectra) and
velocity distribution in the outflow (constant velocity wind) are applied.
They  find that iron  $K_{\alpha}$ photons are generated by the
absorption of X-ray photons at energies higher than the K-edge (i.e $>7$ keV) (see these K-edge features in Fig.3).
Electron scattering of the $K_{\alpha}$ photons within the highly ionized expanding flow leads 
to a decrease of their energy (redshift) which is of first order in $v/c$ (this is clearly illustrated in Fig.1) 
{\it This photon redshift is  an intrinsic property of  any outflow for which divergence is positive}.
LaT04 find  the range of the parameter ${\rm Inner~Radius}/L40$ (which is proportional
to the inverse of the so called ``ionization parameter''
used in the literature) is about $10^{13}$ cm  and 
{\it the range of the wind temperature is about $10^6$ K} 
when the observed $K_{\alpha}$ lines are 
produced in the wind (where $L40$ is the source luminosity in $10^{40}$ ergs s$^{-1}$).
They also find  that the equivalent widths of red skewed Fe K$\alpha$ originated in the wind is order of keV.

\section{CONCLUSIONS}

We have developed an analytic formulation for the
emergent spectrum resulting from photon diffusion in
a  spherically expanding Comptonizing media,
characterized by two model  parameters: an average number of scatterings in the medium $N_{av}$ and 
efficiency the energy loss in the divergent flow $\varepsilon$.
In this formulation of the Radiative Transfer problem, the number of photons 
emitted in the wind shell equals that which escape to the Earth
observer  (i.e. the photon number is conserved). Consequently,  if   the high energy photons lose 
their energy on the way out but the number of
photons is conserved, {\it it  has to result in the accumulation of the photons at a particular lower energy band}.
 
Application of the model to high-energy spectra of
several compact binaries, which have been previously modeled
within the Compton reflection scenarios, seems to lead to
a satisfactory representation of the data. We suggest
that in some instances, the apparent $\sim10-$keV
continuum enhancement seen in Galactic and extra-
galactic sources is due to downscattering effects
associated with an outflow rather than to reflection by
a disk. 

 We may also  conclude {\it that scattering and absorption of the primary line photons in the relatively 
``cold'' outflow (that temperature is few times $10^6$ K) 
 lead to the downscattering modification of the continuum and to 
the formation of red-skewed lines (that is a more natural and probable mechanism than  
the general relativistic effects in the innermost part of the accretion flow)}.

In future work, we will employ detailed Monte-Carlo calculations to
further explore the range of validity. In addition to Galactic
binaries, extra-galactic sources may be characterized by out-flowing
plasma. With the expanding database of INTEGRAL observations, which
will hopefully lead to an increasingly accurate characterization of
the high-energy continua, it may be possible to to explore this
idea. In addition to the continuum, we will explore the possibility of
the line feature formation  in the wind.

\vspace{0.2in}
\centerline {\bf ACKNOWLEDGEMENTS}
    
 L.T.  appreciates  productive  discussions with Philippe Laurent.
L.T. acknowledges the support of this work by Faculty Fellowship Program in NASA
Goddard Space Flight Center. This work made use of the NASA High-Energy Astrophysics
Research Archive Center (HEASARC), and the NASA INTEGRAL
Guest Observer Facility. We also acknowledge the thorough analysis 
of this paper by the referee and his/her constructive and interesting 
suggestions.

\newpage 
\begin{deluxetable}{rrrrrrrrr} 
\tablecolumns{7} 
\tablewidth{0pc} 
\tablecaption{Source Sample and Comptonization Model Parameters}
\tablehead{ 
\colhead{Source ID}
 & \colhead{Observatory} & \colhead{$\Gamma$}  & \colhead{$N_{av}$}   & \colhead{$E_\ast$} 
 & \colhead{$\epsilon$} &  \colhead{Flux} &      \colhead {$\chi^2_\nu$} & \colhead{$DoF$}
}

\startdata 

Cyg X-1 &   INTEGRAL    & 1.74  & 0.8   & 175   & 0.12  & 10.01 & 1.544 & 1359 \\
Cyg X-1 &   RXTE        & 1.51  & 1.7   & 219   & 0.12  & 17.1  & 1.651	& 80 \\
GX339-4 &   RXTE        & 1.63  & 3.2   & 299   & 0.10  & 4.27  & 0.988	& 123 \\
GS1353-64 & RXTE        & 1.35  & 1.9   & 103   & 0.10  & 5.2   & 1.015	& 122 \\
Cyg X-3 &   INTEGRAL    & 2.61  & 1.6   & 160   & 0.10  & 5.57  & 1.218	& 1550 \\
Cyg X-3 &   RXTE        & 2.03  & 1.01  & 270   & 0.10  & 13.1  & 1.068	& 68 \\

\enddata

Flux is for 3-100 keV, units of $10^{-9} \rm \, ergs \, cm^{-2} \, s^{-1}$
\end{deluxetable}

\newpage
\begin{figure}
\includegraphics[width=6in,height=4.in,angle=0]{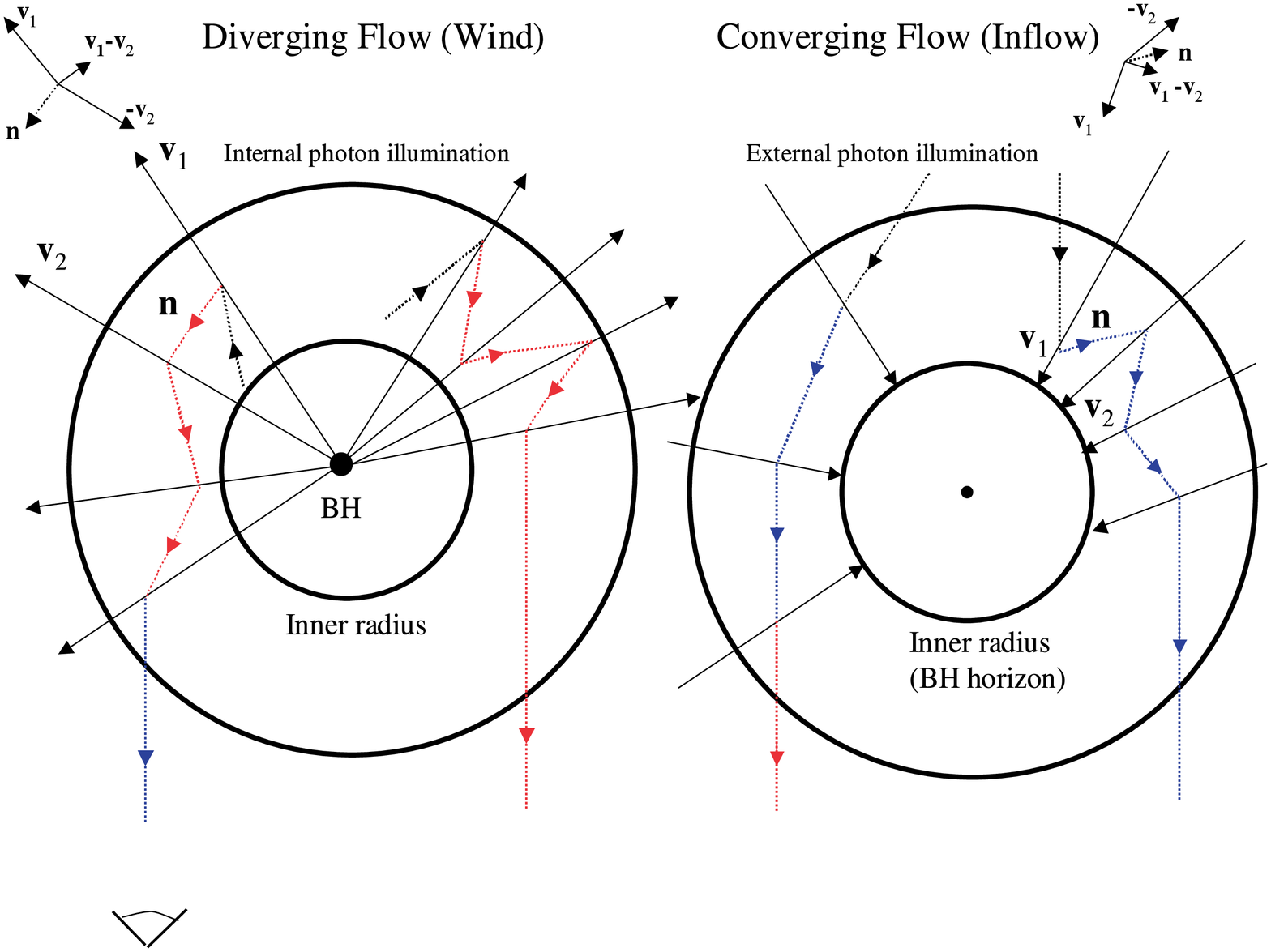}
\caption{On the left side: Schematic diagram depicting the wind 
geometry. The outflow (wind)
originates at the inner radius. The electron optical depth of the wind 
is of order unity.  A photon emitted near the inner boundary and 
subsequently scattered by an electron moving with velocity ${\bf v}_1$, 
impinges on  an electron moving with velocity ${\bf v}_2$ as shown.
The change in
frequency is $\nu _2 = \nu _1\left[1+\left({\bf v}_1-{\bf v}_2\right]
\cdot{\bf n}/c\right)$ where ${\bf n}$ is a unit vector along the
path of the photon at the scattering point.  In a diverging flow
$\left({\bf v}_1-{\bf v}_2\right)\cdot{\bf n}/c <0$ and photons are
successively redshifted, until scattered to an observer at infinity.
 The color of photon path indicates
the frequency shift in the rest frame of the receiver (electron or the Earth
observer). On the right side:  In a converging flow
$\left({\bf v}_1-{\bf v}_2\right)\cdot{\bf n}/c >0$ and photons are
blueshifted.
}
\end{figure}
\newpage 
\begin{figure}
\includegraphics[width=3in,height=3in,angle=0]{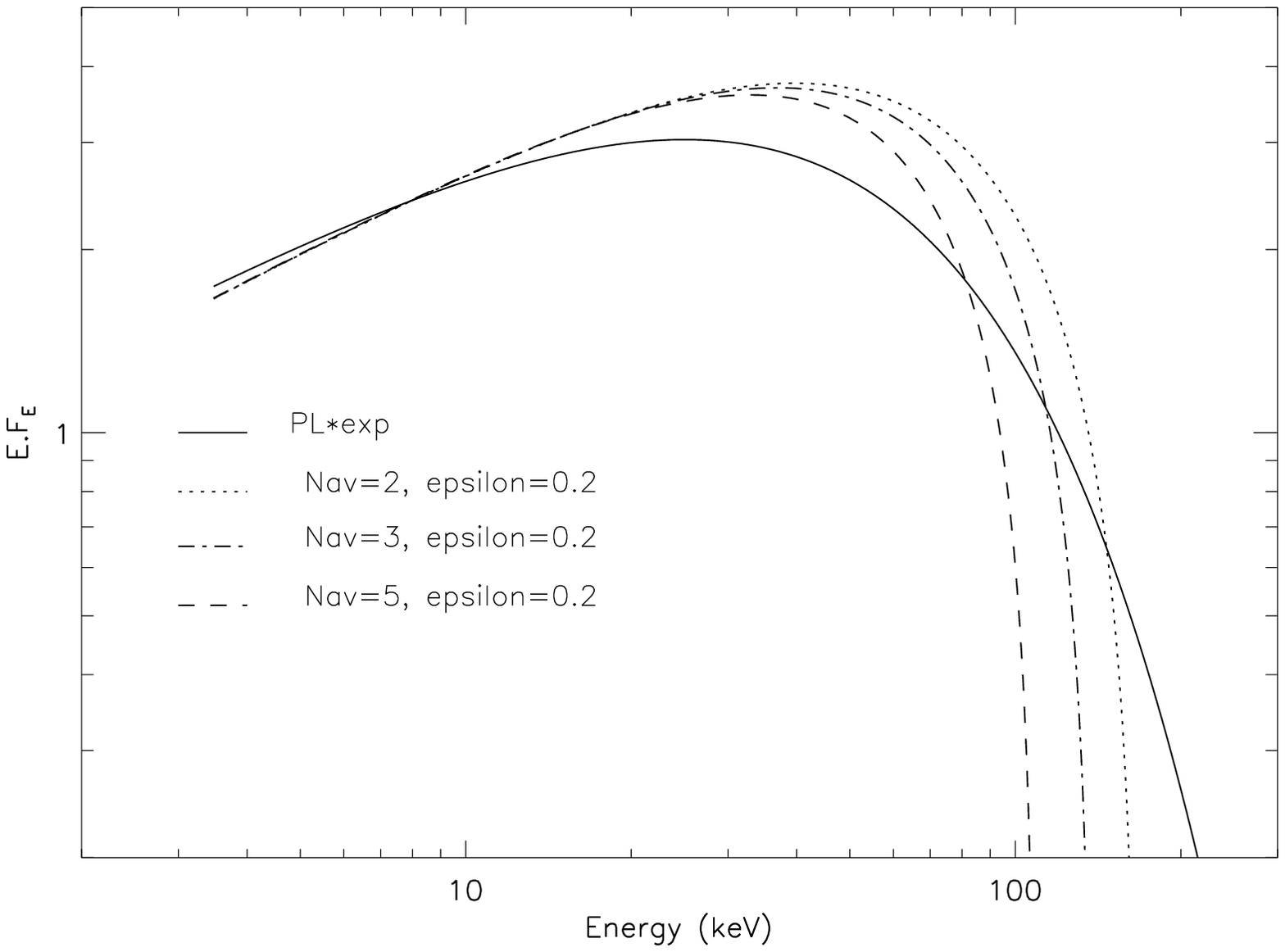}
\includegraphics[width=3in,height=3in,angle=0]{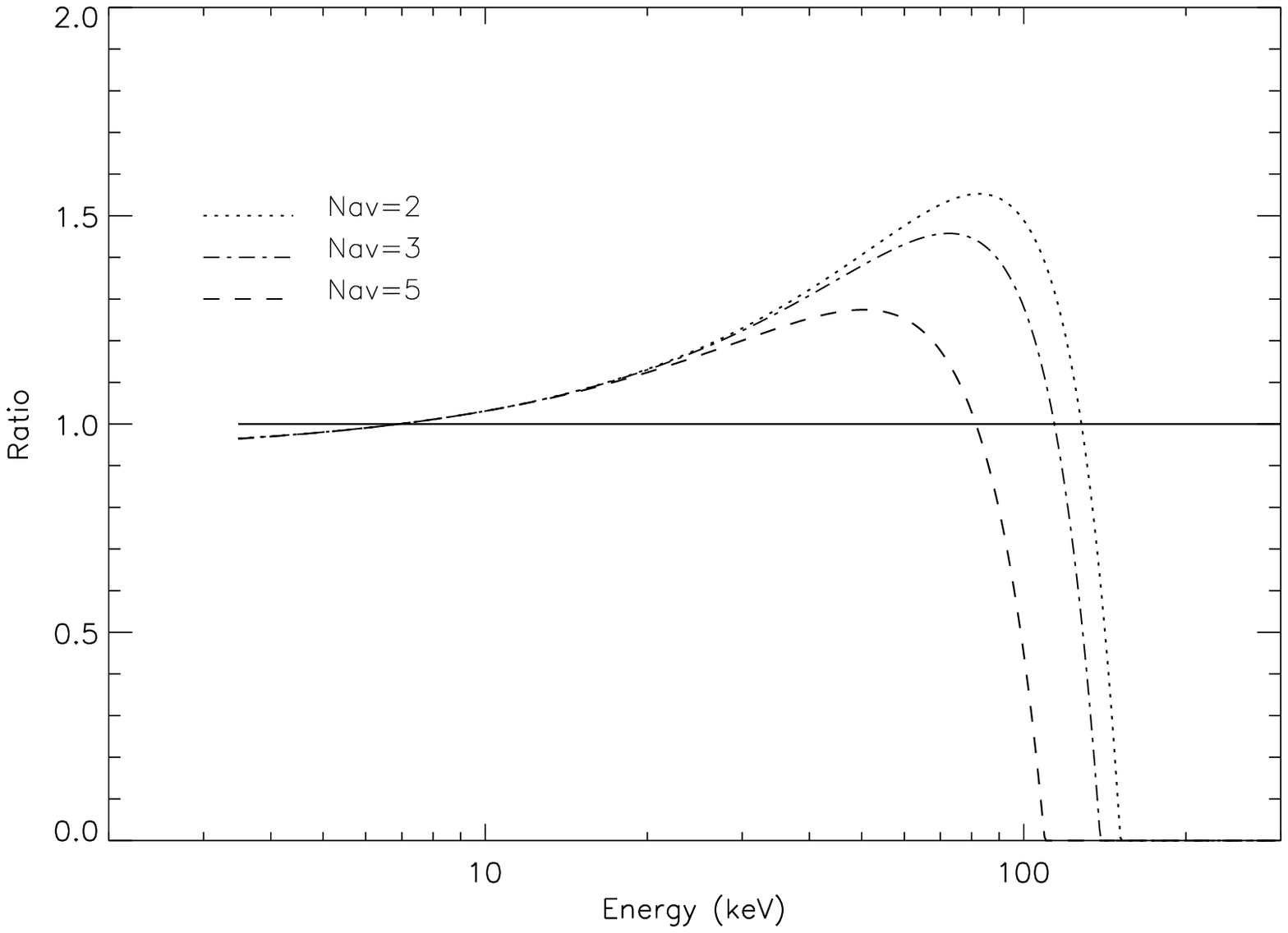}
\caption{a) $E{\cal F}_{E}$ diagram representing the thermal 
Comptonization spectra that result from photon propagation through a divergent
outflow.  The photon source is in the center of the outflow shell.
Solid line: the emission of the central source for which  a spectrum is 
$E^{-\alpha}\exp(-E/E_{\ast})$ where $\alpha=0.5$ and $E_{\ast}=50$ keV. 
Dashed, dot-dashed and dotted lines: the emission
not escaping radiation from the outflow shell for that average number of
scattering $N_{av}=$ 2, ~3,~ 5 respectively. The effective coefficient of the
outflow divergence $\varepsilon=0.2$.
 b) Similar to Figure 2a, except that a ratio of the Comptonization 
models to the exponentially cutoff power-law spectrum is plotted.
}
\end{figure}


\newpage 
\begin{figure}
\includegraphics[width=6.in, height=4in.]{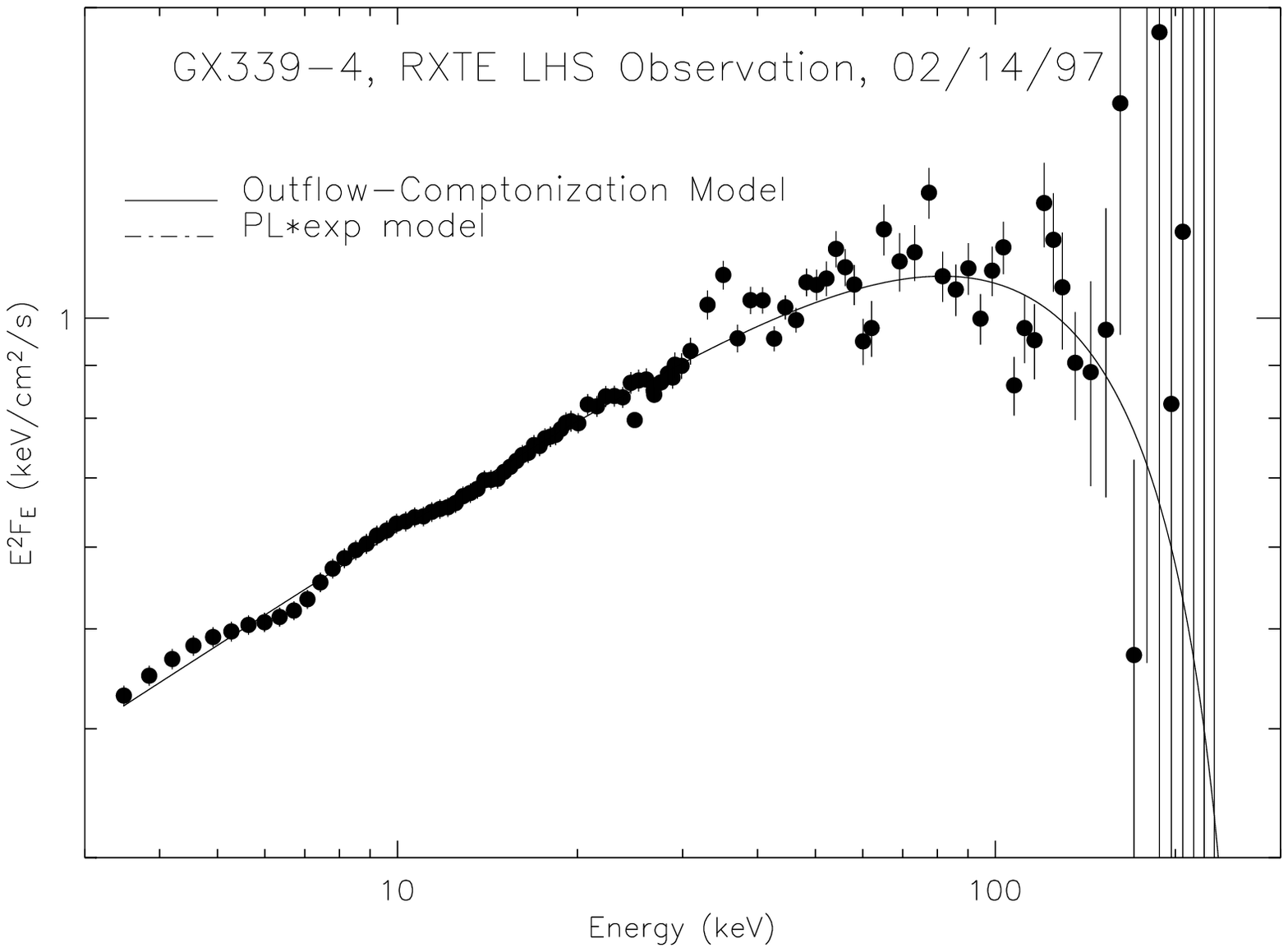}
\includegraphics[width=6.in, height=2.in]{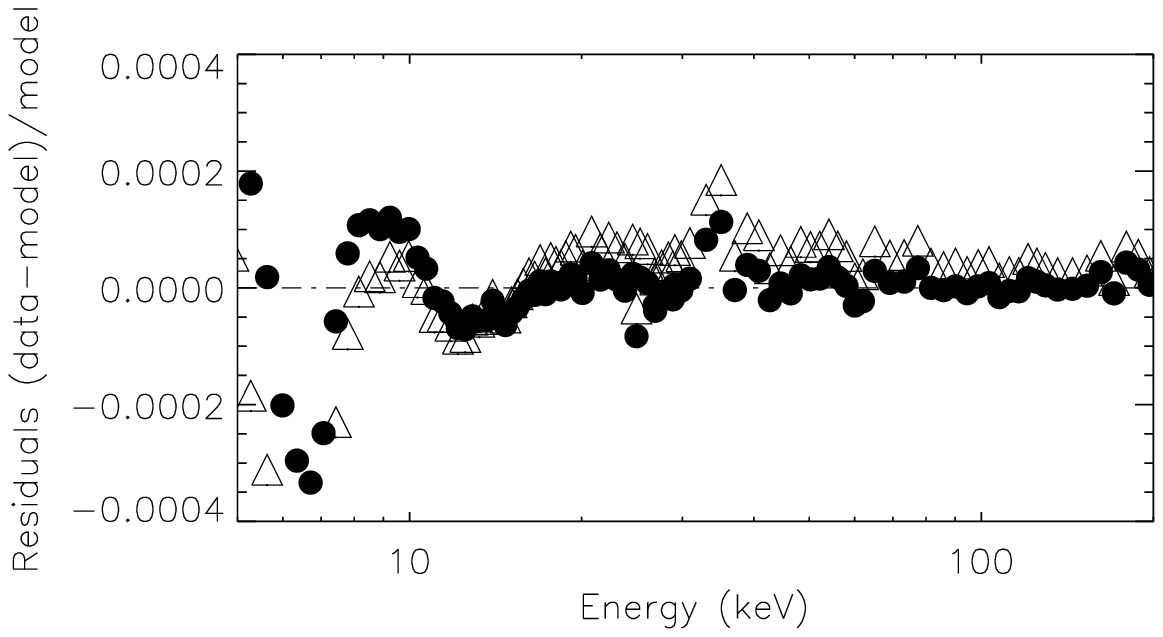}
\caption{a) Comparison between our Comptonizaton model (solid curve), and a 
simple cutoff power-law form, i.e. $\sim KE^{-\Gamma}\exp(-E/E_0)$ (dashed curve). 
Both curves resulted from fits to the overlayed data points, from GX~339-4.
No additional parameters (such as absorption or emission due to Fe) were 
included.
It is evident that the Comptonization model represents the data 
more accurately above $\sim 10 $ keV (the improvement in $\chi^2_\nu$ is
about 10\%). As detailed in the text, we interpret 
the "excess" flux in that regime as being due to downscattering of hard photons
in a diverging outflow. \
b) Fit residuals (for GX 339-4) above 10~keV for the cut-off powerlaw 
model (triangles) and
our out-flow Comptonization model (filled circles). As evidenced from this figure, 
the downscattering effects provide a viable explanation of the 
~10-100~keV "excess" continuum.
}
\end{figure}


\newpage 
\begin{figure}
\includegraphics[width=6in,height=6in,angle=-90]{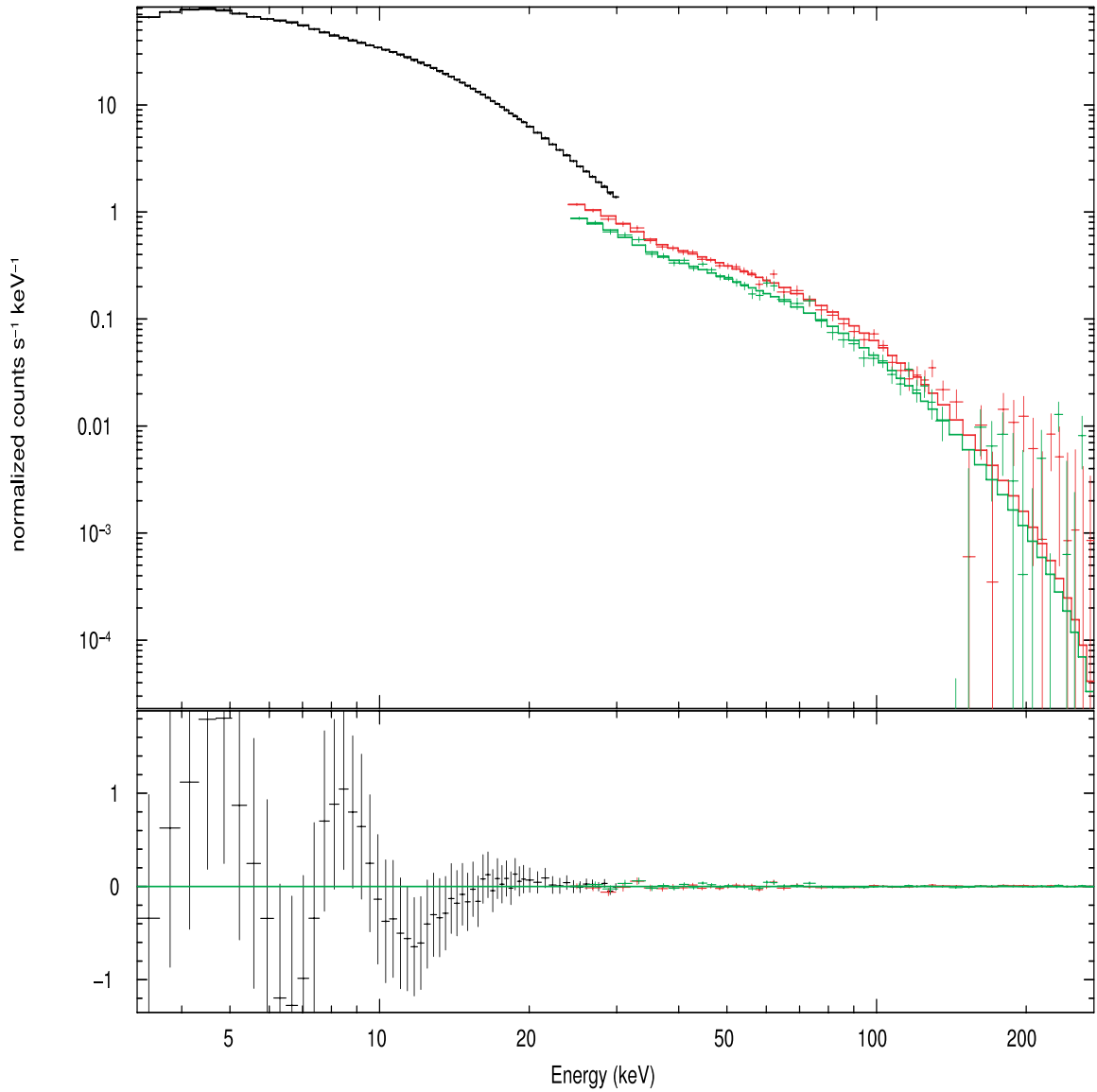}
\caption{ 
Example of the the application of our model to observational data.
In this case, flaring state RXTE data for GS~1351-64 are shown. This 
source has been noted by others as a strong ``Compton reflection'' source among
the Galactic X-ray Binaries. The count rate data,the folded model, and the
residual are plotted. In this case, the chi-square per degree of freedom was
$\chi^2_\nu = 1.02$.
}
\end{figure}

\newpage 
\begin{figure}
\includegraphics[width=6in,height=6in,angle=0]{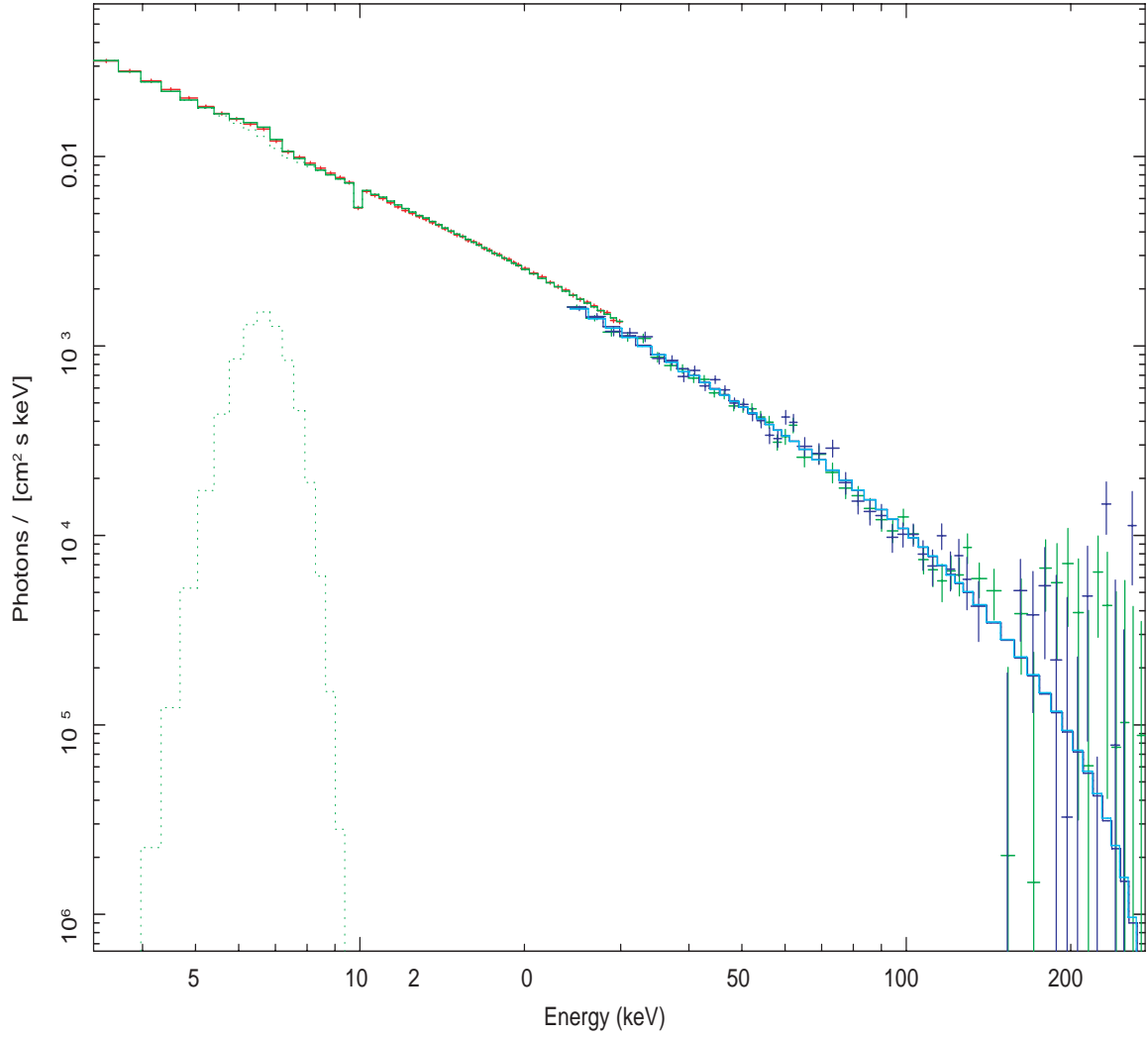}
\caption{ Similar to Figure 4, but the data and model
fit are plotted in photon space.
}
\end{figure}

\begin{thebibliography}{}

\bibitem[Arav, N., 2003]{}
 Arav, N., . 2003, AGN Physics with the Sloan Digital Sky Survey 
 {\it in Proceedings of ASP series, Eds. G.T. Richards and
 P.B. Hall}, astro-ph/0311108

\bibitem[Atti\'e, et al., 2003]{}
Atti\'e, D., et al., 2003, A\&A, 411, L71

\bibitem[Basko et al. (1974)]{BST}
Basko, M.M. \& Sunyaev, R.A. 1974, A\&A, 31, 249 (BST74)

\bibitem[Blandford \& Payne (1981)]{BP81}
Blandford, R.D. \& Payne, D.G. 1981, MNRAS, 194, 1033 (BP81)


\bibitem[Brandt, \& Schulz (2000)]{bs}
Brandt, W.N,, \& Schulz, N.S. 2000, ApJ, 544, 123


\bibitem[Elvis (2003)]{EL}
Elvis, M. 2003, AGN Physics with the Sloan Digital Sky Survey 
 {\it in Proceedings of ASP series, Eds. G.T. Richards and
 P.B. Hall}, astro-ph/0311436


\bibitem[Gilfanov et al. (1999)]{}
Gilfanov, M., et al. 1999, A\&A, 352, 182



\bibitem[Gilfanov et al. (2003)]{}
Gilfanov, M., Churazov, E., Revnivtsev, M., {\it Proceedings 
 of X-Ray Timing 2003: Rossi and Beyond (CfA, 2003), Eds. P.Kaaret, 
 F.K.Lamb, \& J.H.Swank}, astro-ph/0312445

\bibitem[Heindl et al. (2003)]{}
Heindl, W.A.,  et al. 2003,
ApJ,  588,  L97 

\bibitem[Laming \& Titarchuk (2004)]{lt04} 
Laming, L.M.  \& Titarchuk, L. 2004, \apj, submitted (LaT04)

\bibitem[Laurent \& Titarchuk (2004)]{lt04} 
Laurent, P. \& Titarchuk, L. 2004, \apj, submitted (LT04)

\bibitem[Laurent \& Titarchuk (2001)]{lt01} 
Laurent, P. \& Titarchuk, L. 2001, \apj, 562, 67

\bibitem[Laurent \& Titarchuk (1999)]{lt99} 
Laurent, P. \& Titarchuk, L. 1999, \apj, 511, 289



\bibitem[Magdziarz \& Zdziarski (1995)]{}
Magdziarz, P., \& Zdziarski, A. 1995, MNRAS, 273, 837


\bibitem[Markowitz et al.  (2003)]{mev}
Markowitz, A., Edelson, R., \& Vaughan, S. 2003, ApJ, 598, 935


\bibitem[Mattson \& Weaver (2004)] {}
Mattson, B.J., \& Weaver, K.A. 2004, ApJ, 601, 771


\bibitem[Miller et al. (2004)]{mil}
Miller, J. et al.  2004, ApJ, 601, 450 


\bibitem[Nobilli et al. (1993)]{nob93}
Nobili, L. Turolla, R. \& Zampieri, L.  1993, ApJ, 404, 686





\bibitem[Nowak, Wilms \& Dove 2002]{}
Nowak, M.A., Wilms, J., \& Dove, J.B. A\&A, 2002, 332,856

\bibitem[Paizis et al. (2003)]{PA}
Paizis, A. et al.  2003, A\&A, 411, L363

\bibitem[Payne \& Blandford (1981)]{PB81}
Payne, D.G., \& Blandford, R.D. 1981, MNRAS, 196, 781 (PB81)


\bibitem[Pottschmidt et al (2003)]{}
Pottschmidt, K., et al. 2003, A\&A, 471, L383


\bibitem[Proga \& Kallman (2002)]{}
Proga, D., Kallman, T.R. 2002, ApJ, 565, 455

\bibitem[Sheffer, et al., 1992]{}
Sheffer, E.K., et al. 1992, Astron. Zh., 69, 82

\bibitem[Strohmayer \& Brown 2002]{}
Strohmayer, T.E. \& Brown 2002, ApJ, 566, 1045

\bibitem[Sturner, et al., 2003]{}
Sturner, S., et al. 2003, A\&A, 411, L81

\bibitem[Sunyaev \& Titarchuk (1985)]{st85} Sunyaev, R.A. \& 
Titarchuk, L.G. 1980, A\&A, 143, 374 (ST85)

\bibitem[Sunyaev \& Titarchuk (1980)]{st80} Sunyaev, R.A. \& 
Titarchuk, L.G. 1980, A\&A, 86, 121 (ST80)





\bibitem[Titarchuk (1994)]{T94} 
Titarchuk, L. 1994, ApJ, 434, 570

\bibitem[Titarchuk, Kazanas \& Becker (2003)]{TKB} 
Titarchuk, L.,  Kazanas, D. \& Becker, P.A. 2003, ApJ, 598, 411 (TKB03)


\bibitem[Titarchuk, Mastichiadis \& Kylafis (1997)]{TMK97}  
Titarchuk, L.,  Mastichiadis, A. \& Kylafis, N.D. 1997, ApJ, 487, 834 

\bibitem[Titarchuk, et al. (2002)]{TCW} 
Titarchuk, L., Cui, W., \& Wood, P.A. 2002, ApJ, 576, L49 (TCW02)




\bibitem[Vaughan \& Fabian  (2004)]{VF}
Vaughan, S. \& Fabian, A.C. 2004, MNRAS, 348, 1415


\bibitem[Vilhu et al (2003)]{}
Vilhu, O., et al. 2003, A\&A, 471, L405


\end{thebibliography}
\end{document}